\documentclass[11pt]{article}
\usepackage{amsmath}
\usepackage{amsfonts}
\usepackage{amssymb}
\usepackage{graphicx}

\def\a{\alpha}      
\def\b{\beta}

\def\m{\mu} 
\def\o{\omega}

\def\nn{\nonumber}
\def\O{\Omega}

  \def\cR{{\cal R}}

\def \be {\begin{equation}}
\def \ee {\end{equation}}
\def\bea{\begin{eqnarray}}
\def\eea{\end{eqnarray}}

\begin{document}
\begin{center}
\LARGE { \bf Holographic description of  Kerr-Bolt-AdS-dS Spacetimes
  }
\end{center}
\begin{center}
B. Chen$^{**}${ \footnote{bchen01@pku.edu.cn}},
A. M. Ghezelbash$^*${ \footnote{masoud.ghezelbash@usask.ca}},  V. Kamali$^\dagger$\footnote{vkamali1362@gmail.com},
M. R. Setare$^\dagger$\footnote{rezakord@ipm.ir}
\\
$^{**}$Department of Physics,
and State Key Laboratory of Nuclear Physics and Technology,
and Center for High Energy Physics,
Peking University,\\
Beijing 100871, P.R. China\\
$^*$Department of Physics and Engineering Physics, \\ University of Saskatchewan, \\
Saskatoon, Saskatchewan S7N 5E2, Canada\\
$^\dagger$ Department of Science of Bijar, Kurdistan University  \\
Bijar, Iran
\end{center}
\vskip 1cm

\begin{abstract}
We show that there exists a holographic 2D CFT description of a Kerr-Bolt-AdS-dS spacetime. We first consider the wave equation of a massless scalar field propagating in extremal Kerr-Bolt-AdS-dS spacetimes and find in the "near region", the wave equation in extremal limit could be written in terms
of the $SL(2,R)$ quadratic Casimir. This suggests that there exist dual CFT descriptions
of these black holes. 
In the probe limit, we compute the
scattering amplitudes of the scalar off the extremal black holes and find perfect agreement
with the CFT prediction. Furthermore we study the holographic description of the generic four dimensional
non-extremal Kerr-Bolt-AdS-dS black holes. We find that if focusing on
the near-horizon region, for the massless scalar scattering in the low-frequency limit, the radial equation could still be rewritten as the $SL(2,R)$ quadratic Casimir, suggesting the existence of dual 2D description. We read the temperatures of the dual CFT from the conformal coordinates and obtain the central charges by studying the near-horizon geometry of near-extremal black holes. We recover the macroscopic entropy from the microscopic counting.  We also show that for the superradiant scattering, the retarded Green's functions
and the corresponding absorption cross sections are in perfect match with CFT prediction.

\end{abstract}

\newpage

\section{Introduction}

The Kerr/CFT correspondence states that the Kerr black hole could be described holographically by a two-dimensional conformal field theory(CFT), even though the Kerr 
spacetime is asymptotically flat. It 
was first proposed for extremal Kerr black hole in \cite{1} by studying its near-horizon geometry  (NHEK) \cite{2}, and was
then improved by the subsequent study \cite{3}. The main progress were made essentially on the extremal and near extremal limits in which the black hole near horizon geometries consist a certain AdS structure and the central charges of dual CFT can be obtained by analyzing
the asymptotic symmetry following the method in \cite{4} or by calculating the boundary stress tensor of the $2D$ effective action \cite{5}.
One support of this conjecture has been found in the perfect match of the macroscopic Berenstein-Hawking entropy of the black hole with the conformal field theory entropy computed via the Cardy formula. Another support of the correspondence came from the 
study of the superradiant scattering off the black hole. It was shown in \cite{Bredberg:2009pv}
that the bulk scattering amplitudes
were in precise agreement with the CFT descriptions whose form are
completely fixed by the conformal invariance.
Similar discussions have been generalized to charged
Kerr-Newman \cite{Hartman:2009nz}, multi-charged
Kerr \cite{Cvetic:2009jn} and higher dimensional near-extremal Kerr
black holes. In \cite{ChenChu}, it was shown that the real-time correlators of
various perturbations in near-extremal Kerr(-Newman) black hole
could be computed directly from the bulk, following the Minkowski
prescription proposed in AdS/CFT \cite{Son05}. 
The similar
prescription has been applied to calculate the three-point
functions in the Kerr/CFT correspondence \cite{Becker:2010jj}. 
 
Recently, the Kerr/CFT correspondence was
generalized to generic non-extremal Kerr black hole \cite{61}. In paper \cite{61} the authors argued that the
existence of conformal invariance in a near horizon geometry is not a necessary condition,
instead the existence of a local conformal invariance in the solution space of the wave
equation for the propagating field is sufficient to ensure a dual CFT description. The
observation indicates that even though the near-horizon geometry of a generic Kerr black
hole could be far from the AdS or warped AdS spacetime, the local conformal symmetry on
the solution space may still allow us to associate a CFT description to a Kerr black hole.
Both the microscopic entropy counting and the low frequency scalar scattering amplitude in
the near region support the picture. The study of hidden conformal
symmetry has been generalized to other kinds of black hole \cite{71} and applied to compute the
real-time correlators in Kerr/CFT \cite{Chen:2010xu}.

Even though the hidden conformal symmetry has been applied to the study of holographic descriptions of various black holes, its origin is still a puzzle. Indeed, for the various black holes in 3D and 4D  which have holographic description, there exist such hidden conformal symmetry, in match with the fact that the dual CFTs have the same symmetry. In \cite{61}, it was
argued that the symmetry comes from the freedom in choosing the matching region between the ``far" and ``near" regions. However, the study on the holographic description of a Kerr(-Newman)-AdS-dS black hole \cite{Chen:2010bh} suggests that this may not be the case. For the Kerr-AdS-dS black hole, the hidden conformal symmetry acting on the solution space make sense only in the restricted near horizon region, indicating that the hidden conformal symmetry could be related to the universal behavior of the black hole. 

In this paper, we would like to discuss the holographic description of Kerr-Bolt-AdS-dS black holes. The extreme Kerr-Sen and Kerr-Bolt black holes have been studied in \cite{GH2} and \cite{8}, following the treatment in \cite{1}. For the generic Kerr-Bolt black hole, its hidden conformal symmetry and holographic description has been discussed in \cite{7}. For a  Kerr-Bolt-AdS-dS black hole, due to the presence of cosmological constants, the treatment is trickier. 

We start from the extremal black hole by analyzing the hidden conformal symmetry of radial equation. From the conformal coordinates introduced in \cite{6}, we read the temperatures of the dual CFT. The central charge could be obtained from the near horizon geometry of the black hole. This provides a holographic picture of the extremal black hole. Both the microscopic counting and the study of the low-frequency scattering amplitude support this picture. 

Furthermore, we generalize the study to the generic non-extremal case, following the treatment in \cite{Chen:2010bh}. In this case, only in the near horizon region, the radial equation could be written in terms of $SL(2,R)$ quadratic Casimir. In this procedure, the 
breaking of the conformal symmetry by the periodic identification on angular coordinate allows us to read the left and right temperatures of the dual CFT. The central charges are computed for the near-extremal black hole and are assumed to be true even for generic case. The microscopic entropy via the Cardy formula reproduce exactly the Bekenstein-Hawking entropy. The other evidence is from the study of the superradiant scattering, whose scattering amplitude is in perfect agreement with the CFT prediction.

\section{Kerr-Bolt-AdS-dS black hole }

In this section, we give a brief review of the Kerr-Bolt-AdS-dS black hole ( \cite{8}, \cite{7}). The metric of Kerr-Bolt-dS takes the following form in terms of Boyer-Lindquist type coordinates
\begin{eqnarray}
ds^2&=&-\frac{\Delta_r}{\Xi^2\rho^2}[dt+(2n\cos\theta -a\sin^2\theta)d\varphi]^2+\frac{\Delta_{\theta}\sin^2\theta}{\Xi^2\rho^2}[adt-(r^2+n^2+a^2)d\varphi]^2
\nonumber\\
&+&
\frac{\rho^2dr^2}{\Delta_r}+\frac{\rho^2d\theta^2}{\Delta_\theta},
\label{metric1}
\end{eqnarray}
where
\begin{eqnarray}
\rho^2&=&r^2+(n+a\cos\theta)^2,~~~~~~~~~~~~~~~~~~~~~~~~~~~~~~~~~~~~~~~~~~~~~~~~\nn\\
\Delta_r&=&-\frac{r^2(r^2+6n^2+a^2)}{l^2}+r^2-2mr-\frac{(3n^2-l^2)(a^2-n^2)}{l^2}, \nonumber\\
\Delta_{\theta}&=&1+\frac{a\cos\theta(4n+a\cos\theta)}{l^2},~~~~~~~~~~~~~~~~~~~~~~~~~~~~~~~~~~~~~~~\nonumber\\
\Xi&=&1+\frac{a^2}{l^2}.~~~~~~~~~~~~~~~~~~~~~~~~~~~~~~~~~~~~~~~~~~~~~~~~~~~~~~~~~~~~~~~
\label{Dr}
\end{eqnarray}
Here $n$ is the NUT charge. The event horizons of the black hole are given by the singularities of the metric function which are the real roots of $\Delta_r=0$.
 The rang of $\theta$ depends strongly on the values of the NUT charge $n$, the rotational
 parameter $a$ and the cosmological constant $\Lambda =\frac{3}{l^2}$. For the Kerr-Bolt-AdS with negative cosmological constant $l^2$ should be replaced with $-l^2$ in metric (\ref{metric1}) and metric functions (\ref{Dr}).
 
The angular velocity and the Bekenstein-Hawking entropy are respectively
 \be
 \Omega_H=\frac{a}{r^2_++n^2+a^2}, \hspace{3ex}S=\frac{\pi(r^2_++n^2+a^2)}{\Xi}.
\ee

The surface gravity of the cosmological horizon can be calculated to
give%
\begin{equation}
\kappa =\frac{1}{2(r_{+}^{2}+n^{2}+a^{2})\Xi }\left. \frac{d\Delta _r }{d r }
\right| _{r=r_{+}}
\end{equation}%
where the Killing vector 
$\chi ^{\mu }=\zeta ^{\mu }+\Omega_H \psi^{\mu}$
is normal to the horizon surface $r=r_{+}$.

We first note that there are no pure NUT solutions for nonzero values of $a$. 
Since $\psi ^{\mu }$\ is a Killing vector,
for any constant $\varphi $ -slice near the horizon, additional conical
singularities will be introduced in the $(t,r)$\ section unless the period
of $t$\ is $\Delta t=\frac{2\pi }{\left| \kappa \right| }$. Furthermore,
there are string-like singularities along the $\theta =0$ and $\theta =\pi $
axes for general values of the parameters. \ These can be removed by making
distinct shifts of the coordinate $t$ in the $\varphi $ direction near each of
these locations. These must be geometrically compatible 
\cite{GHKB},
yielding the requirement that the period of $t$ should be
$\Delta t=4n\Delta \phi=\frac{16\pi n}{q_++q_-}$. Demanding the absence of both
conical and Dirac-Misner singularities, we get the relation 
\begin{equation}
  \frac{k}{\kappa}=\frac{8n}{q_++q_-} 
  \label{betaeq8Pin}
\end{equation}%
where $k$ is any non-zero positive integer and $q_+$ cand $q_-$ are two relatively prime integers. Demanding the existence of a
pure NUT solution at $r=r_{+}$ is equivalent to the requirement that 
the area of the surface of the fixed point set of the Killing vector $%
\partial /\partial t,$ vanishes. 
In other words, this surface is of
zero dimension. This can only occur special values of the mass parameter.
However if we select for this parameter we find an inconsistency with the
relation (\ref{betaeq8Pin}), which must hold for the spacetime with NUT
charge.

Hence we conclude that the only spacetimes with NUT charge and rotation is
simply Kerr-Bolt-AdS/dS, where the term ``bolt'' refers to the fact
that the dimensionality of the fixed point set of $\partial /\partial t$
is two.
%
%

Now we consider a bulk massless scalar field $\Phi$ propagating in the background of (\ref{metric1}). The Klein- Gordon(KG) equation
\begin{eqnarray}\label{6}
\Box\Phi=\frac{1}{\sqrt{-g}}\partial_{\mu}(\sqrt{-g}g^{\mu\nu}\partial_{\nu})\Phi=0
\end{eqnarray}
 can be simplified by assuming the following form of the scalar field
\begin{eqnarray}\label{7}
  \Phi(t,r,\theta,\varphi)=\exp(-i\omega t+i m\varphi)S(\theta)  R(r)
\end{eqnarray}
where $\omega$ and $m$ are the quantum numbers corresponding to the translational symmetry along $t$ and $\varphi$. The wave equation reduces to two decoupled equations by separation of variables
\begin{eqnarray}
\partial_r(\Delta\partial_r R)+[\frac{\Xi^2((r^2+n^2+a^2)\omega-ma)^2}{\Delta_r}+2\Xi^2ma\omega-\lambda]R&=&0,\label{5}\\
\frac{1}{\sin\theta}\partial_\theta(\sin\theta\partial_\theta S)+(\lambda-f(\theta))S&=&0~~~~~~~~~~~\label{66}
\end{eqnarray}
where $\lambda$ is the separation constant and
\begin{eqnarray}\label{77}
f(\theta)=\frac{\Xi^2}{\Delta_{\theta}}(\frac{m^2}{\sin^2\theta}+\frac{(2n\cos\theta-a\sin^2\theta)^2\omega^2}{\sin^2\theta}+2m\omega a\cos\theta(4n+a\cos\theta)).
 \end{eqnarray}

Let us first consider the extreme Kerr-Bolt black hole. In this case the
Hawking temperature is vanishing, $\Delta=(r-r_+)^2$. In the low
frequency limit, and in the near region which is defined by
\begin{eqnarray}
 r\ll\frac{1}{\omega}~~~~~~~~~~~~~~~~M\ll\frac{1}{\omega}~~~~~~~~~~~~~~~~~~~n\ll\frac{1}{\omega}, \nn \end{eqnarray}
 we have the radial equation 
\begin{eqnarray}
\left\{\partial_r(\Delta\partial_r)+\frac{\Xi^2}{K^2}\frac{4r_+\omega ((r_+^2+a^2+n^2)\omega-ma)}{r-r_+}\right.~~~~~~~~~~~~~~~~~~~~~~\nn\\
\left.+\frac{\Xi^2}{K^2}\frac{((r_+^2+a^2+n^2)\omega-ma)^2}{(r-r_+)^2}\right\}R(r)=\lambda
R(r)~~~~~~~~~~~~~~\label{radialex}
\end{eqnarray}
where 
\be
K=1-\frac{6r_+^2}{l^2}-\frac{6n^2+a^2}{l^2}.
\ee

\section{Holographic description of extreme Kerr-Bolt-AdS-dS}

In this section, we show that there exist holographic 2D CFT description of extreme Kerr-Bolt-AdS-dS. In the literature\cite{1,{Hartman:2008pb}}, the standard treatment is to focus on the near horizon geometry of the extremal black hole and analyze the asymptotic symmetry group to get the central charge of the dual CFT, while the temperature of the dual CFT could be read from the Hartle-Hawking vacuum. This kind of treatment has  been applied to the study of the extremal Kerr-Bolt  in \cite{8}.  Here we take a slightly different approach to find the holographic dual. By studying the radial equation (\ref{radialex}), we find the hidden conformal symmetry acting on the solution space. This suggests that there exists a 2D CFT dual. Moreover, this provides a simple way to read the temperature in the dual CFT. However, we still need the near-horizon geometry to read the central charges. 

Let us start from the conformal coordinates introduced in \cite{6}
\begin{eqnarray}
\omega^{+}&=&\frac{1}{2}(\alpha_1 t+\beta_1 \varphi-\frac{\gamma_1}{r-r_+}),~~~~~~\label{11}
\\
\omega^{-}&=&\frac{1}{2}(\exp(2\pi T_L\varphi+2n_Lt)-\frac{2}{\gamma_1}), \label{12}\\
 y&=&\sqrt{\frac{\gamma_1}{2(r-r_+)}}\exp(\pi T_L\varphi+n_L t), \label{13}
\end{eqnarray}
from which  we can define two sets of  vector fields
\begin{eqnarray}
H_1&=&i\partial_{+},~~~~\nn\\
H_0&=&i(\omega^{+}\partial_{+}+\frac{1}{2}y\partial_{y}),~~~~\label{H}\\
~H_{-1}&=& i((\omega^{+})^2\partial_{+}+\omega^{+}y\partial_{y}-y^2\partial_{-})~~~\nn
\end{eqnarray}
and
\begin{eqnarray}
\overline{H}_1&=&i\partial_{-}, \nn\\
\overline{H}_0&=&i(\omega^{-}\partial_{-}+\frac{1}{2}y\partial_{y}), \label{overH}\\
\overline{H}_{-1}&=&i((\omega^{-})^2\partial_{-}+\omega^{-}y\partial_{y}-y^2\partial_{+}). \nn
\end{eqnarray}
Each set of the vector fields satisfies the commutation relation of the $SL(2,R)$ algebra. \begin{eqnarray}\label{20}
 ~~[H_0,H_{\pm1}]=\mp i H_{\pm 1},~~~~~~~~[H_{-1},H_1]=-2iH_0
\end{eqnarray}
and
\begin{eqnarray}\label{21}
~[\overline{H}_0,\overline{H}_{\pm1}]=\mp i \overline{H}_{\pm 1},~~~~~~~~[\overline{H}_{-1},\overline{H}_1]=-2i\overline{H}_0. 
\end{eqnarray}

The quadratic $SL(2,R)$ Casimir is
\begin{eqnarray}\label{22}
H^2=\widetilde{H}^2=-H_{0}^2+\frac{1}{2}(H_1H_{-1}+H_{-1}H_{1})=\frac{1}{4}(y^2\partial_{y}^2-y\partial_{y})+y^2\partial_{+}\partial_{-}, \end{eqnarray}
which could be rewritten 
   in term of $ \varphi,t$ and $r$ as 
\begin{eqnarray}
H^2=\partial_r(\Delta\partial_r)-(\frac{\gamma_1(2\pi T_L\partial_t-2n_L\partial_{\varphi})}{A(r-r_+)})^2~~~~~~~~~~~~~~~~\nn\\
-\frac{2\gamma_1(2\pi T_L\partial_t-2n_L\partial_{\varphi})}{A(r-r_+)}(\beta_1\partial_t-\alpha_1\partial_{\phi})~~~~~~~~~~~~~~\label{Casimirex}
\end{eqnarray}
where $A=2\pi T_L\alpha_1-2n_L\beta_1$ and $\Delta=(r-r_+)^2$. The crucial observation is that
the equation (\ref{radialex}) could  be rewritten as the $SL(2,R)$ Casimir (\ref{Casimirex}) 
\begin{eqnarray}\label{31}
H^2\Phi(r)=\lambda\Phi(r).
\end{eqnarray}
with the following identification
\begin{eqnarray}\label{24}
&&\alpha_1=0,~~~~~~~~~~~\beta_1=\frac{\gamma_1}{a},~~~~~~~~T_L=\frac{K}{\Xi}\frac{r_+^2+a^2+n^2}{4\pi
a r_+}, ~~~~~~~~~~n_L=-\frac{\Xi}{4r_+ K}.\nonumber\\
&&
\end{eqnarray}

The existence of the hidden conformal symmetry suggests that extreme Kerr-Bolt-AdS-dS black hole could be described by a dual ``chiral"  CFT with a nonvanishing left temperature and a vanishing right temperature. 

To obtain the central charges, we have to zoom in the near horizon of the extremal black hole. For our purpose, we consider the near horizon geometry of the near-extremal black hole. Similar to the Kerr case, we have the following coordinate transformation
\begin{eqnarray}
r=\frac{r_+ +r_*}{2}+\epsilon r_0 \hat{r},~~~~~~r_+-r_*=\mu\epsilon r_0~~~~~~t=\frac{\Xi r_0}{\epsilon}\hat{t}\nonumber\\
\varphi=\varphi+\frac{ar_0}{\epsilon(r_+^2+a^2+n^2)}\hat{t}~~~~~~~~~~~~~~~~~~~~~~~~~~~\label{NEHKBcoor}
\end{eqnarray}
then we have the near-horizon geometry of near-extremal Kerr-Bolt-AdS-dS  black hole
\begin{eqnarray}
ds^2&=&\Gamma(\theta)(-(\hat{r}-\frac{\mu}{2})(\hat{r}+\frac{\mu}{2})
d\hat{t}^2+\frac{d\hat{r}^2}{(\hat{r}-\frac{\mu}{2})(\hat{r}+\frac{\mu}{2})}+\alpha(\theta)d\theta^2)
\nonumber\\
&+&\gamma(\theta)(d\hat{\varphi}+\tilde{p}\hat r d\hat{t})^2~~~~~~~~~~~~~~~~~~~~~~~~~~~~~~~~~~~~~~~~~~~~~~~~~~~~\label{Near-NHEKB}
\end{eqnarray}
 where
\begin{eqnarray}\nn
\Gamma(\theta)=\frac{\rho_+^2r_0^2}{r_+^2+a^2+n^2},~~~~~~\alpha(\theta)=\frac{r_+^2+a^2+n^2}{\Delta_\theta r_0^2},~~~~~~\gamma(\theta)=\frac{\Delta_\theta\sin^2\theta(r_+^2+a^2+n^2)^2}{\rho_+^2r_0^2\Xi^2}~~~~~\\
\nonumber
\tilde{p}=\frac{ar_0^2\Xi(r_+ +r_*)}{(r_+^2+a^2+n^2)^2},~~~~~~\rho_+^2=r_+^2+(n+a\cos\theta)^2,~~~~~~~r_0^2=\frac{r_+^2+a^2+n^2}{K}.~~~~~~~~~~~~~~~~
\end{eqnarray}

For the extremal black hole, $r_+=r_\ast$. The central charge could be obtained from the analysis of the asymptotic symmetry group. It turns out to be just the following integral
\be
c_L=3\tilde{p}\int_0^\pi d\theta\sqrt{\Gamma(\theta)\alpha{(\theta)}\gamma{(\theta)}}=12\frac{ar_+}{K}.
\ee

The first evidence for this holographic picture comes from the counting of the microscopic degrees of freedom. In the extremal limit, the microscopic entropy just comes from the left sector and for Kerr-Bolt-AdS-dS black holes 
\begin{eqnarray}\label{43}
S=\frac{\pi^2}{3}c_LT_L=\frac{\pi(r_+^2+a^2+n^2)}{\Xi}
\end{eqnarray}
in agreement with macroscopic Bekenstein-Hawking entropy.

In the case of vanishing cosmological constant $l^{-2}=0$, we have the Kerr-Bolt black hole with $\Xi=0, K=1$ and $r^2_+=a^2-n^2$ such that 
\be
T_L=\frac{a}{2\pi r_+}, \hspace{3ex}c_L=12a r_+.
\ee 
This is exactly the result found in \cite{8}. 

 We can determine the conjugate  charge from the first low of thermodynamics.
 We identify
\begin{eqnarray}\label{466}
\delta M=\omega,~~~~~~~~~~~~~~~\delta E_L=\omega_L
\end{eqnarray}
with
\begin{eqnarray}\label{477}
\omega_L=\frac{(r_+^2+a^2+n^2)}{a}\omega
\end{eqnarray}
then
\begin{eqnarray}\label{488}
\delta S=\frac{\delta E_L}{T_L}=\frac{\omega_L}{T_L}
\end{eqnarray}

The radial equation can be written as
\begin{eqnarray}\label{32}
[\partial_r(\Delta\partial_r)+\frac{B}{r-r_+}
+\frac{C^2}{(r-r_+)^2}]R(r)=\lambda R(r)~~~~~~~~~~~~~~
\end{eqnarray}
where
\begin{eqnarray}\label{33}
C&=&\frac{\Xi}{K}((r_+^2+a^2+n^2)\omega-ma)~~~~~~~~~~~\\
B&=&\frac{\Xi^2}{K^2}4r_+\omega((r_+^2+a^2+n^2)\omega-ma)
\end{eqnarray}

Introducing $z=\frac{-2iC}{r-r_+}$, we get the equation
\begin{eqnarray}\label{34}
\frac{d^2R}{dz^2}+(\frac{\frac{1}{4}-m_s^2}{z^2}+\frac{k}{2}-\frac{1}{4})R(z)=0
\end{eqnarray}
where
\begin{eqnarray}\label{35}
k=i\frac{2r_+\omega\Xi}{K},~~~~~~~~~~~~~m_s^2=\frac{1}{4}+\lambda
\end{eqnarray}
This equation has the solution
\begin{eqnarray}\label{36}
R(z)=C_1R_+(z)+C_2R_-(z)
\end{eqnarray}
where
\begin{eqnarray}\label{37}
R_{\pm}(z)=\exp(-\frac{z}{2})z^{\frac{1}{2}\pm m_s}F(\frac{1}{2}\pm m_s-k,1\pm2m_s,z)
\end{eqnarray}
are two linearly independent solution. Here $F$ is the Kummer function and could be expand in two limits. In the near horizon region $r\rightarrow r_+$ so $z\rightarrow \infty$ and 
\begin{eqnarray}\label{38}
F(\alpha.\gamma,z)\sim\frac{\Gamma(\gamma)}{\Gamma(\alpha-\gamma)}\exp(-i\alpha\pi)z^{-\alpha}+\frac{\Gamma(\gamma)}{\Gamma(\alpha)}\exp(z)z^{\alpha-\gamma}.
\end{eqnarray}
 On the other hand, when $r$ goes asymptotically to infinity, $z\rightarrow 0$, $F\rightarrow 1$. 
 
 We need to impose the purely ingoing boundary condition at $r=r_+$. This requires that 
 \begin{eqnarray}\label{41}
C_1=-\frac{\Gamma(1-2m_s)}{\Gamma(\frac{1}{2}-m_s-k)}C_0,~~~~~~~~~~C_2=\frac{\Gamma(1+2m_s)}{\Gamma(\frac{1}{2}+m_s-k)}C_0,
\end{eqnarray}
with $C_0$ being a constant. Asymptotically 
 the solution behaves as
\begin{eqnarray}\label{39}
R\sim C_1r^{-h}+C_{2}r^{1-h}
\end{eqnarray}
where $h$ is the conformal weight
\begin{eqnarray}\label{400}
h=\frac{1}{2}+m_{s}=\frac{1}{2}+\sqrt{\frac{1}{4}+\lambda}. 
\end{eqnarray}

The retarded Green function could be read directly \cite{ChenChu,{Chen:2010xu}} 
\begin{eqnarray}
G_R\sim\frac{C_1}{C_2}\propto\frac{\Gamma(1-2h)\Gamma(h-k)}{\Gamma(2h-1)\Gamma(1-h-k)}.
\end{eqnarray}
With the identification on $\omega_L$ and $T_L$, 
\begin{eqnarray}\nn
G_R&\sim&\frac{\Gamma(1-2h)\Gamma(h-i\frac{2r_+\omega\Xi}{K})}{\Gamma(2h-1)\Gamma(1-h-i\frac{2r_+\omega\Xi}{K})}\\
&=& \frac{\Gamma(1-2h)\Gamma(h-i\frac{\omega_L}{2\pi
T_L})}{\Gamma(2h-1)\Gamma(1-h-i\frac{\omega_L}{2\pi T_L})}.\label{Grex}
\end{eqnarray}
The real-time correlator (\ref{Grex}) is obviously in agreement with the CFT prediction. This provides another support to the holographic picture.

\section{Holographic description of generic Kerr-Bolt-AdS-dS black hole  }

 For Kerr-Bolt black hole (with $\Lambda=0$), if we focus on near region, the radial equation should be rewritten in terms of the $SL(2,R)$ quadratic Casimir \cite{7}. The same treatment does not work in the case of non-extremal Kerr-Bolt-AdS-dS black hole as the function $\Delta_r$ is quartic. However, the situation is quite similar to the Kerr-Newman-AdS-dS case. Following the treatment in \cite{8},  we expand the function $\Delta_r$  in the near horizon region to the quadratic order of $(r-r_+)$,
\begin{eqnarray}\label{8}
\Delta_r=-\frac{r^2(r^2+6n^2+a^2)}{l^2}+r^2-2mr-\frac{(3n^2-l^2)(a^2-n^2)}{l^2}\\
\nonumber\simeq K(r-r_+)(r-r_*)~~~~~~~~~~~~~~~~~~~~~~~~~~~~~~~~~~~~~~~~~~~~~
  \end{eqnarray}
where $r_{+}$ is the outer horizon, and
\begin{eqnarray}
K&=&1-\frac{6r_+^2}{l^2}-\frac{6n^2+a^2}{l^2},~~~~~~~~~~~~~~~~~~~~~~~~~~~~~~~~~~~\label{9}\\
r_*&=&r_+ +\frac{1}{K}(\frac{3r_+^3}{l^2}+(\frac{6n^2+a^2-l^2}{l^2})r_+ +\frac{(3n^2-l^2)(n^2-a^2)}{l^2r_+}).\label{10}
\end{eqnarray}
In general, $r_{*}$ is not the inner horizon. Only in the case that $\Delta_r$  is quadratic, which happens in the Kerr-Bolt which zero cosmological constant case, $r_{*}$ coincides with the other horizon. In the low frequency limit and in near horizon region the radial equation should be simplified even more
\begin{eqnarray}
\partial_r(r-r_+)(r-r_*)\partial_r R(r)+\left\{\frac{\Xi^2(ma-\omega(r_+^2+a^2+n^2))^2}{K^2(r-r_+)(r_+ -r_*)}\right.\nn\\
\left. -\frac{\Xi^2(ma-\omega(r_-^2+a^2+n^2))^2}{K^2(r-r_*)(r_+ -r_*)}\right\}R(r)=\frac{\lambda}{K}R(r)~~~~~~~~~~~~~~\label{radialNE}
\end{eqnarray}


Following \cite 6 we now show that the equation (\ref{radialNE}) can be reproduced by the introduction of conformal coordinates.
 We will show that for massless scalar field $\Phi$, there exist a hidden $SL(2,R)_{L}\times SL(2,R)_{R}$ conformal symmetry acting on the solution space. Moreover, from the spontaneous breaking of this hidden symmetry by periodic identification of $\varphi$, we can read out the left and right temperature of dual conformal field theory.
 We introduce the conformal coordinates
\begin{eqnarray}
\omega^{+}&=&\sqrt{\frac{r-r_{+}}{r-r_{-}}}\exp(2\pi T_{R}\varphi+2n_{R}t)~~\\
\omega^{-}&=&\sqrt{\frac{r-r_{+}}{r-r_{-}}}\exp(2\pi T_{L}\varphi+2n_{L}t) \\
 y&=&\sqrt{\frac{r_{+}-r_{-}}{r-r_{-}}}\exp(\pi( T_{R}+T_{L})\varphi+(n_{R}+n_{l})t)~~~
\end{eqnarray}
Similarly, from this set of coordinates, one can locally define two sets of $SL(2,R)$ vector fields as (\ref{H}) and (\ref{overH}).  The key point is that the equation (\ref{radialNE}) can be rewritten as the quadratic $SL(2,R)$ Casimir with the following identification:
\begin{equation}
 T_{R}=K\frac{r_+ -r_*}{4\pi a\Xi}, ~~~~~~T_{L}=K\frac{r_+^2+r_*^2+2(n^2+a^2)}{4\pi a(r_+ +r_-)\Xi},~~~\label{temperature}
\end{equation}
\begin{equation}
~~~~~~~n_{L}=-\frac{\Xi}{2(r_+ +r_-)K},~~~~~~~~~~~~~~~~~~~~~n_R=0.~~~~~~~~~~~~~~~~~
\end{equation}

>From the identification, so we know the corresponding left and right temperatures in the dual CFT. In order to have a microscopic description of the black hole we need to determine the central charge of the dual CFT. In Kerr case, it was conjectured that the central charges in the near extremal limit should be generalized to the generic cases \cite{6}. This treatment has been proved to be valid in the Kerr-Newman, Kerr-Newman-AdS-dS black holes \cite{Chen:2010bh}. The near-horizon geometry of the near extreme black hole is of the form (\ref{Near-NHEKB}). 
As in \cite{Hartman:2008pb,{Chen:2010bh}}, the central charge should be
\begin{eqnarray}\label{centralcharge}
c_R=c_L=3\tilde{p}\int_0^\pi d\theta\sqrt{\Gamma(\theta)\alpha{(\theta)}\gamma{(\theta)}}=6a\frac{r_+ +r_*}{K}.
\end{eqnarray}
Here we assume that the left and right central charges should be same, similar to the Kerr case. When we take $l^{-2}=0$, the above central charges reduce to the ones in the Kerr-Bolt case \cite{7}. 

The macroscopic Bekenstein-Hawking entropy of the black hole could be recovered from the microscopic counting in the dual CFT via Cardy formula
\begin{eqnarray}\label{399}
S_{CFT}=\frac{\pi^2}3({c_{L}T_{L}+c_{R}T_{R}})~
\end{eqnarray}
>From the central charges (\ref{centralcharge}) and temperatures (\ref{temperature}) and using equation (\ref{399}) we have
\begin{eqnarray}\label{40}
S=\frac{\pi(r_+^2+a^2+n^2)}{\Xi}, 
\end{eqnarray}
which is exactly the Bekenstein-Hawking entropy of a generic Kerr-Bolt-AdS-dS black hole. 

The further support to the holographic picture comes from the study of the superradiant scattering. From the radial equation of the scalar scatering off the 
 Kerr-Bolt-AdS-dS black hole background, we see that the scattering issue may not well-defined.  In general, the radial equation is hard to treat as it involves four singularity. However as we showed before the equation looks tractable in the very near region of the horizon. Unfortunately, as we move a little away from the horizon, the expansion does not make sense and we cannot find the asymptotical behavior of the scalar eigenfunction in a consistent way. Nevertheless for the near-extremal black hole, if we 
 focus on the superradiant scattering, the issue is well-defined. This is similar to the 
  Kerr-Newman-AdS-dS case. 
  
 The near-horizon geometry of the near-extremal Kerr-Bolt-AdS-dS black hole has been 
 given in (\ref{Near-NHEKB}) with rescaled coordinates (\ref{NEHKBcoor}). Due to the scaling on the coordinates $(t,\phi)$, we have to focus on the frequencies very close to the super-radiant bound
 \be
 \o-m\O_H=\hat \o \frac{\epsilon}{r_0}. 
 \ee
In this case, the angular equation is frequency-independent and the radial equation is 
just 
\bea
[\partial_{\hat r}(\hat r-\frac{\mu}{2})(\hat r+\frac{\mu}{2})\partial_{\hat r}+\frac{A_s}{\hat r-\frac{\mu}{2}}+\frac{B_s}{\hat r+\frac{\mu}{2}}+C_s]R=0~~~~
\eea
with
\begin{eqnarray}
A_s&=&\frac{\Xi^2\hat \o}{\mu}~~~~~~~~~~~~~~~~~~~~~~~~~~~~~\nn\\
B_s&=&-\frac{\Xi^2\mu}{K^2}\left(\frac{\hat \o}{\mu}-\frac{2m\O_H}{K}r_+\right)^2~~~~~~~~~~~~~~~~~~~~~~~~~~~\nn
\end{eqnarray}
and $C_s$ is the separation constant, determined by the angular equation. 
The wavefunction satisfying the ingoing boundary condition at the horizon is of the form
\be
\cR(z)=z^\a(1-z)^\b F(a_s,b_s,c_s;z)
\ee
where $z=\frac{\hat r-\mu/2}{\hat r+\m/2}$
\begin{eqnarray}\label{44}
\alpha=\sqrt{A_s},~~~~~~~~~~~~~~\beta=\frac{1}{2}(1-\sqrt{1-4C_s}),~~~~~~~~~~~~~~~\gamma=\sqrt{-B_s}
\end{eqnarray}
and
\begin{eqnarray}\label{45}
c_s=1-2i\alpha,~~~~~~~~~~~~~a_s=\beta+i(\gamma-\alpha),~~~~~~~~~~~~~b_s=\beta-i(\gamma+\alpha)
\end{eqnarray}

The asymptotic form can be read out by taking the limit $z\rightarrow 1$ and $1-z\rightarrow r^{-1}$, 
\begin{eqnarray}\label{46}
R(\hat r\to \infty)\sim D r^{h_s-1}+Er^{-h_s}
\end{eqnarray}
where
\begin{eqnarray}\label{47}
D=\frac{\Gamma(c_s)\Gamma(2h_s-1)}{\Gamma(a_s)\Gamma(b_s)},~~~~~~~~~~~~~~~~~~~E=\frac{\Gamma(c_s)\Gamma(1-2h_s)}{\Gamma(c_s-a_s)\Gamma(c_s-b_s)},
\end{eqnarray}
and $h_s$ is the conformal weight 
\be
h_s=\frac{1}{2}(1+\sqrt{1-4C_s}). 
\ee
The essential properties of the absorption cross section is indeed captured by the coefficient $D$ such as
\begin{eqnarray}\label{48}
P_{abs}\sim\mid D\mid^{-2}\sim\sinh(2\pi\alpha)\mid\Gamma(a_s)\mid^2\mid\Gamma(b_s)\mid^2
\end{eqnarray}
Alternatively, we can read the real-time correlator directly\cite{ChenChu,{Chen:2010xu}}
\be
G_R\sim \frac{E}{D}=\frac{\Gamma(1-2h_s)}{\Gamma(2h_s-1)}\frac{\Gamma(a_s)\Gamma(b_s)}{\Gamma(c_s-a_s)\Gamma(c_s-b_s)},
\ee
from which we can read the absorption cross section easily $P_{abs}=Im(G_R)$.

To see explicitly that $P_{abs}$ matches with the microscopic grey body factor of the dual CFT, we needs to identify the conjugate charges, $\delta E_L$ and $\delta E_{R}$ defined by
\begin{eqnarray}\label{49}
\delta S_{BH}=\delta S_{CFT}=\frac{\delta E_L}{T_L}+\frac{\delta E_R}{T_R}
\end{eqnarray}
From the first law of black hole thermodynamics
\begin{eqnarray}\label{50}
T_H\delta S_{BH}=\delta M-\Omega_H\delta J
\end{eqnarray}
 the solution is
\begin{eqnarray}\label{51}
\delta E_L&=&\frac{(r_+^2+r_-^2+2(a^2+n^2))}{2a\Xi}\delta M,\\
\label{52}
\delta E_R&=&\frac{(r_+^2+r_-^2+2(a^2+n^2))}{2a\Xi}\delta M-\delta J
\end{eqnarray}
If we identify
\begin{eqnarray}\label{53}
\delta M=\omega,~~~~~~~~~~~~~~~\delta J=m
\end{eqnarray}
\begin{eqnarray}\label{54}
\omega_L=\frac{(r_+^2+r_*^2+2(a^2+n^2))}{2a\Xi}\omega,~~~~\omega_R==\frac{(r_+^2+r_*^2+2(a^2+n^2))}{2a\Xi}\omega-m, 
\end{eqnarray}
we have
\begin{eqnarray}\label{55}
\delta E_L=\omega_L,~~~~~~~~~~~~~~~~~\delta E_R=\omega_R
\end{eqnarray}

For the superradiant scattering, it is easy to see that 
 the coefficients $a_s$ and $b_s$ can be expressed in terms of parameters $\omega_L$ and $\omega_R$
\begin{eqnarray}\label{57}
a_s=h_R+i\frac{\omega_R}{2\pi T_R},~~~~~~~~~~~~~~~~~~~~~~b_s=h_L+i\frac{\omega_L}{2\pi T_L}
\end{eqnarray}
where for the scalar $h_L=h_R=h_s$. 
Similarly we have 
\begin{eqnarray}\label{58}
2\pi \alpha=\frac{\omega_L}{2 T_L}+\frac{\omega_R}{2 T_R}.
\end{eqnarray}
Finally from (\ref{48}), (\ref{57}) and (\ref{58}) the absorption cross section can be expressed as
\begin{eqnarray}\label{433}
P_{abs}\sim T_{L}^{2h_{L}-1}T_{R}^{2h_{R}-1}\sinh(\frac{\omega_{L}}{2T_{L}}+\frac{\omega_{R}}{2T_{R}})\mid\Gamma(h_{L}+i\frac{\omega_{L}}{2\pi T_{L}})\mid^2\\
\nonumber
 \times\mid\Gamma(h_{R}+i\frac{\omega_{R}}{2\pi T_{R}})\mid^2~~~~~~~~~~~~~~~~~~~~~~~~~~~~~~~~~~~~~~~~~~~~~~~
\end{eqnarray}
which is the finite temperature absorption cross section for $2D$ CFT.\\

\section{Conclusion}

In this paper, we showed that there exist a holographic 2D CFT description for a Kerr-Bolt-AdS-dS black hole. In our study, the key ingredient is the hidden conformal symmetry in the black hole. For the extremal case, the hidden symmetry acting on the solution space of the radial equation in the low frequency limit and in the near region. 
While for the generic non-extremal case, the hidden conformal symmetry only make sense 
in the very near horizon region. This fact is in accordance with the universal property of the black hole. Moreover, it brings another interesting question: in which situation we 
can say a black hole has a holographic description? Our study suggests that even in the near horizon region if the radial equation turns to have a hidden conformal symmetry, there may exist a dual 2D CFT description. 

\section*{Acknowledgments}

 CB was  supported in part by NSFC Grant
No.10775002,10975005. AMG was supported in part by the Natural Sciences and Engineering Research Council of Canada (NSERC).

\end{document}